\documentclass[manuscript]{acmart}

\AtBeginDocument{%
  \providecommand\BibTeX{{%
    \normalfont B\kern-0.5em{\scshape i\kern-0.25em b}\kern-0.8em\TeX}}}

\copyrightyear{2023}
\acmYear{2023}
\setcopyright{rightsretained}
\acmConference[TEI '23]{TEI '23: Proceedings of the Seventeenth International Conference on Tangible, Embedded, and Embodied Interaction}{February 26-March 1, 2023}{Warsaw, Poland}
\acmBooktitle{TEI '23: Proceedings of the Seventeenth International Conference on Tangible, Embedded, and Embodied Interaction (TEI '23), February 26-March 1, 2023, Warsaw, Poland}\acmDOI{10.1145/3569009.3573120}
\acmISBN{978-1-4503-9977-7/23/02}

\begin{document}

\title{CompuMat: A Computational Composite Material for Tangible Interaction}

\author{Xinyi Yang}
\email{xinyiyang@gsd.harvard.edu}
\affiliation{%
  \institution{Harvard University}
  \city{Cambridge}
  \state{MA}
  \country{USA}
}

\author{Martin Nisser}
\affiliation{%
 \institution{Massachusetts Institute of Technology}
 \city{Cambridge}
 \state{MA}
 \country{USA}}
 
\author{Stefanie Mueller}
\affiliation{%
 \institution{Massachusetts Institute of Technology}
 \city{Cambridge}
 \state{MA}
 \country{USA}}

\renewcommand{\shortauthors}{Xinyi Yang, et al.}


\begin{abstract}
  This paper introduces a computational composite material comprising layers for actuation, computation and energy storage. Key to its design is inexpensive materials assembled from traditionally available fabrication machines to support the rapid exploration of applications from computational composites. The actuation layer is a soft magnetic sheet that is programmed to either bond, repel, or remain agnostic to other areas of the sheet. The computation layer is a flexible PCB made from copper-clad kapton engraved by a fiber laser, powered by a third energy-storage layer comprised of 0.4mm-thin lithium polymer batteries. We present the material layup and an accompanying digital fabrication process enabling users to rapidly prototype their own untethered, interactive and tangible prototypes. The material is low-profile, inexpensive, and fully untethered, capable of being used for a variety of applications in HCI and robotics including structural origami and proprioception.

\end{abstract}

\ccsdesc[500]{Human-centered computing~Human computer interaction (HCI)}

\keywords{Digital fabrication, Flexible electronics, Rapid prototyping}


\maketitle

\begin{figure}[H]
  \includegraphics[width=\textwidth]{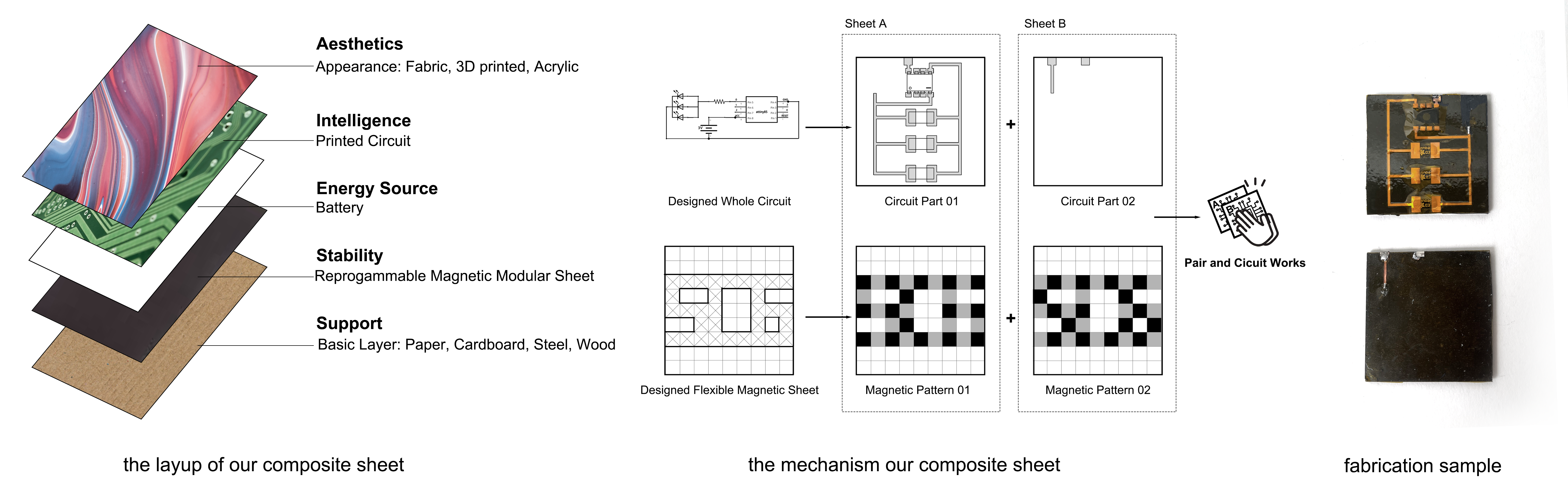}
  \centering
  \caption{\textbf{Layup} The layup of our composite sheet includes a structural support layer, a programmable magnetic sheet that selectively bonds to other sheets, a flexible PCB layer and a 0.4mm-thin lithium polymer battery layer to power it, and an optional layer to convey desired affordances or aesthetics. \textbf{Mechanism overview} A circuit is divided into two parts, and bonded to two magnetic sheets coded with opposite magnetic patterns. The sheets will only bond in one configuration, closing the circuit. \textbf{Applications} A fabrication sample shows a composite square with exposed pads for electronic components can be soldered to achieve specific applications when mated to a complementary square.}
  \label{fig: fig1}
\end{figure}

\section{Introduction \& Overview}

Since the vision of Digital Bits proposed in Ishii’s 1997 paper~\cite{ishii1997tangible}, tangible user interfaces have been explored in a wide range of forms. In exploring how to transition from interacting with bits on desktop
digital devices to interacting with digital bits embedded in physical objects, researchers have explored creating objects that controlled digitally using electromagnets~\cite{nisser2021programmable} or that shape-change predictably via linearly actuated pistons~\cite{follmer2013inform} or soluble support  material~\cite{nisser2019sequential}. Moving to increase portability and inspired in many ways by papercraft, growing communities of researchers turned to fabricating interactive material sheets. This work is strongly informed by seminal research on proposed Computational Composites~\cite{vallgaarda2007computational} or Robotic Materials~\cite{mcevoy2015materials} that combine sensing, actuation and computation in single composite material. For example, Printscreen~\cite{olberding2014printscreen} developed thin film sheets embedded with both capacitive touch and display capabilities and Foldio~\cite{olberding2015foldio} built on this work to enable self-foldability via embedded material actuators. Such laminate composite materials capable of both sensing and actuation have more recently been developed using actively controlled shape memory polymer~\cite{nisser2016feedback}. As these actuators suffer from power-inefficiencies and required offboard energy storage, more recent techniques have leveraged programmable magnetic sheets to enable both physical actuation and
interaction~\cite{nisser2022mixels}. However, these new materials have proven challenging to manufacture as composite sheets capable of sensing, computation, actuation and energy storage in an untethered, portable manner. While foundational work has been done to design materials fitting these criteria, an under-explored area has been the design of such a computational composite that leverages inexpensive, commercially available materials that can be manufactured using traditionally available fabrication machines. 

In this paper, we design such a composite from readily available materials and manufacturing platforms with the aim to accelerate experimentation and prototyping of applications for computational composite materials. We leverage sheets of programmable magnetic sheet to build on prior technologies and propose an inexpensive material layup (Figure \ref{fig: fig1}) that can be fabricated using typical makerspace machines (Figure \ref{fig: pipeline}) that combines sensing, actuation and energy storage for general purpose use. We use the Mixels interface~\cite{nisser2022mixels} to produce pairs of magnetic sheet that attract or repel in one configuration, but
remain agnostic in all others. The desired functionality is given to the accompanying UI, which produces the required
magnetic pixel patterns and plots them using a magnetic plotter. We then use a fiber laser to rapidly engrave custom
PCB pairs whose circuits close electrically only when superimposed in their correct configuration. By bonding these
PCB pairs to magnetic sheet pairs and battery layers, we produce a magnetically actuated electronic sheet that only
bonds in a particular, self-aligning orientation, which closes its circuit to become electronically functional (Figure \ref{fig: pipeline}). We
use the sheet for applications including tangible interactions and origami-inspired self-folding.

\section{Background and Related Work}
\par 
In this section, we outline related work in smart laminate materials and methods that have leveraged magnetism to imbue objects with interactive capabilities.

\subsection{Computational Composite Materials}

Computational Composites~\cite{vallgaarda2007computational} or Robotic Materials~\cite{mcevoy2015materials} that combine sensing, actuation and computation in single composite material, or portable sheets capable of interacting with the user, have been explored by researchers across a variety of communities in the form of tangible devices, papercraft, and foldable robots. Researchers have used printed circuits to create lightweight and portable interfaces that harvest energy from human interaction~\cite{karagozler2014paper}, enable paper-based computing through interactive pop-up books~\cite{qi2010electronic,zhao2017instangible} and that leverage thin film displays to create paper-based sheets capable of dynamic visual feedback for educational tool-kits~\cite{klamka2017illumipaper}.

By adhering together structural layers, embedded electronics, actuators and flexural hinges, roboticists have developed robots that can be manufactured as smart laminate sheets before self-folding into their target geometries. Using paper and carbon fiber as structural backing to flexible PCBs, researchers have embedded actuators in the form of shape memory alloys~\cite{firouzeh2015robogami}, shape memory polymers~\cite{nisser2016feedback}, and pneumatically activated polymer pouches~\cite{niiyama2015pouch}. However these actuators typically suffer from power-inefficiencies and require offboard energy from a dedicated supply to actuate them, reducing portability.

In this work, we substitute away energy-intensive material actuators in favor of reprogrammable magnetic materials capable of exerting near-field forces without any online power. These materials are instead programmed to exert particular forces before use, simplifying force control, and reducing both the cost and thickness of the sheet.

\subsection{Magnetically actuated interactive objects}


Magnets have been widely used to create actuated interfaces by manipulating the magnetic fields surrounding an object. Researchers have used digitally controlled electromagnets to move objects across desks~\cite{pangaro2002actuated}, fabricated paper-based media that enable both movement and dynamic interaction\cite{ogata2015fluxpaper}, utilized magnets to develop VR interaction tool-kits \cite{zhao2017instangible}, and invented portable displays by embedding shielded magnetic designs~\cite{liang2014gaussstones}. In recent years, researchers have explored the feasibility of using magnetic sheet by programming patterns onto them for both haptic and other novel interactive applications~\cite{yasu2017magnetic,yasu2019magnetact}. In particular, Mixels~\cite{nisser2022mixels} built upon work on orthonormal magnetic codes~\cite{nisser2022selective} to develop a magnetic plotter and accompanying user interface that could be used to plot magnetic patterns that were selectively attractive and agnostic to other patterns. 

In this paper, we build on Mixels by integrating their sheet into a material stack of PCBs and battery layers to develop a portable smart laminate material, and develop a digital fabrication pipeline to support its creation. The programmed magnetic sheets allows the material stack to bond selectively to only particular parts of other sheets, creating opportunities in snappable PCBs and error-correcting folding of origami-inspired structures.

\section{Design and fabrication pipeline for programmable composite materials}

This section introduces the design and fabrication procedure for making magnetically programmable composite materials. We begin by introducing the material layup and detail the software and hardware involved in fabricating the finished laminate.

\subsection{Material layup}


Our smart composite material is comprised of three key layers to enable actuation, computation and energy storage, respectively, and costs approximately \$6 USD per 50mm square.

\textbf{Actuation layer} The actuation layer is a soft magnetic sheet that is programmed using the Mixels interface to either bond, repel, or remain agnostic to other areas of the sheet. We use flexible, inexpensive off-the-shelf fridge magnet that is traditionally used as stickers refrigerators and white boards. Thicker sheet is capable of supporting stronger magnetic attraction, while thinner sheet is more lightweight and amenable to bending. We sampled four material thicknesses (0.1mm, 0.55mm, 0.76mm and 1mm) of generic magnetic sheet and chose 0.76mm sheet as a trade-off between these desirable criteria.

\textbf{Computation layer} The computation layer is a flexible PCB made from copper-clad kapton engraved by a fiber laser. We use sheet with a copper trace thickness of 0.3mm which gives high conductivity with insignificant voltage drops and retaining sufficient rigidity to prevent breakage during handling, while remaining flexible enough to allow rolling of the sheet. Copper traces permit electronic components to be soldered directly onto the copper pads during fabrication.

\textbf{Energy storage layer} The computation layer is powered by a third, energy-storage layer, which is comprised of a 0.4mm-thin flexible lithium polymer battery (Powerstream). 
These batteries are rechargeable, flexible and exhibit an operating voltage up to 3.6V which supports interfacing with COTS electronic components.

\textbf{Optional layers} Two additional, optional layers introduced in Figure \ref{fig: fig1} are designated as a structural layer, to give further support to the sheet for load-bearing applications, and an aesthetic layer, to give the sheet a desired texture or color; in future work, this could be augmented with thin film displays~\cite{olberding2014printscreen}.
 




\begin{figure}
  \includegraphics[width=\textwidth]{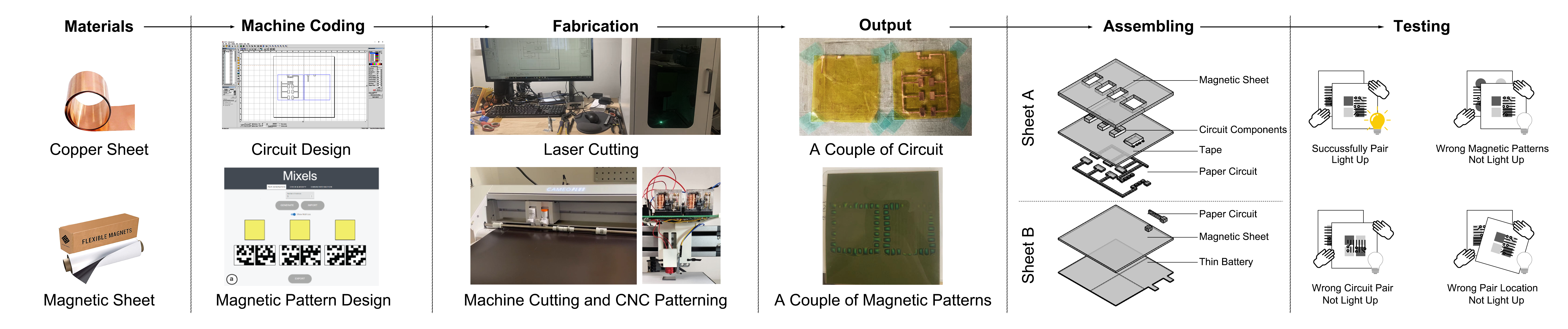}
  \caption{Design and fabrication pipeline: We use copper-clad kapton and soft magnetic sheet to encode with circuits and magnetic codes, respectively; We design circuits and code the magnetic patterns using our own user interface, outputting dxf files for the fiber laser and magnetic plotter to cut and plot; These produce pairs of flexible PCBs and magnetic sheets with desired patterns; We populate the PCBs with components and assemble the material layups into sheet A and sheet B; The sheet becomes electrically functional when sheets A and B are properly aligned, in which orientation they bond magnetically.}
  \label{fig: pipeline}
\end{figure}

\begin{figure}
  \includegraphics[width=\linewidth]{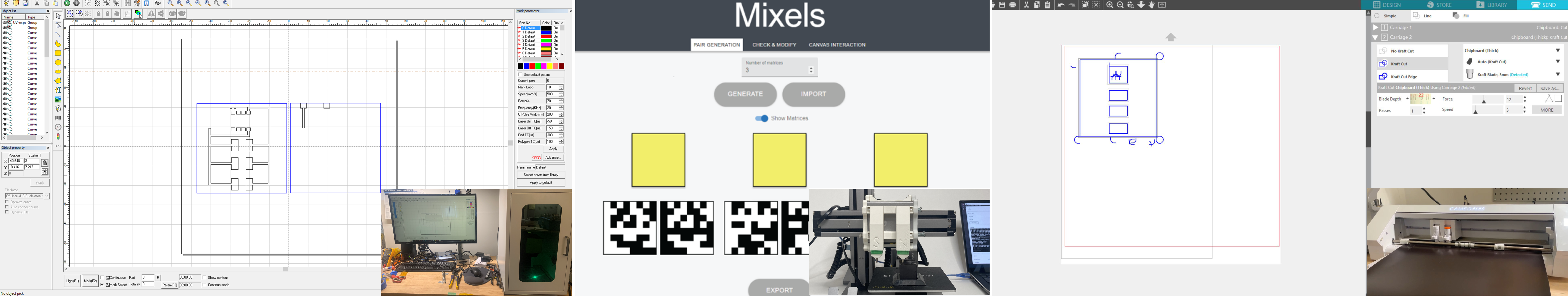}
  \caption{Designing (left) circuits for the fiber laser, (middle) magnetic programs using the Mixels user interface, and (right) magnetic sheets for cutting on a vinyl cutter.}
  \label{fig: User Interface}
\end{figure}

\subsection{Fabrication pipeline}
The design and fabrication pipeline for fabricating our smart composite material is illustrated in Figure \ref{fig: pipeline}. These produce pairs of flexible PCBs and magnetic sheets with desired patterns; We populate the PCBs with components and assemble the material layups into sheet A and sheet B; The sheet becomes electrically functional when sheets A and B are properly aligned, in which orientation they bond magnetically.

\textbf{Materials}: The process begins with selecting sheets of the copper-clad kapton and soft magnetic sheet described in the materials section above.

\textbf{Design}:  We design our circuit pad layouts in Adobe Illustrator and output dxf files for the fiber laser software (zCAD) to execute on the copper-clad kapton (Figure \ref{fig: User Interface}). We design square magnetic sheet geometries for simplicity and use the Mixels user interface to design selectively attractive patterns to plot onto these sheets using the magnetic plotter. By convention we refer to two mating magnetic sheets as A and B, where Mixels ensures that A and B will only bond attractively in one configuration while remaining magnetically agnostic in all other translations and rotations.

\textbf{Fabrication}: We execute the circuit pattern on the copper clad kapton using the fiber laser, upload a desired interface design into the vinyl cutter software (Silhouette Studio) to cut desired sizes of magnetic sheet (for simplicity, we use squares), and encode the magnetic programs onto the soft magnetic sheets using the Mixels magnetic plotter. This plotter consists of a Snapmaker 3-in-1 augmented with an add-on supporting an Arduino Nano micro-controller, an electromagnet, an H-bridge, and a hall effect sensor encased in a 3D-printed housing.

\textbf{Assembly}: After soldering any required electronic components onto the finished PCB, we adhere the PCB layer and a battery layer onto each magnetically programmed sheet (A and B) using CA glue.

\begin{figure}
  \includegraphics[width=\linewidth]{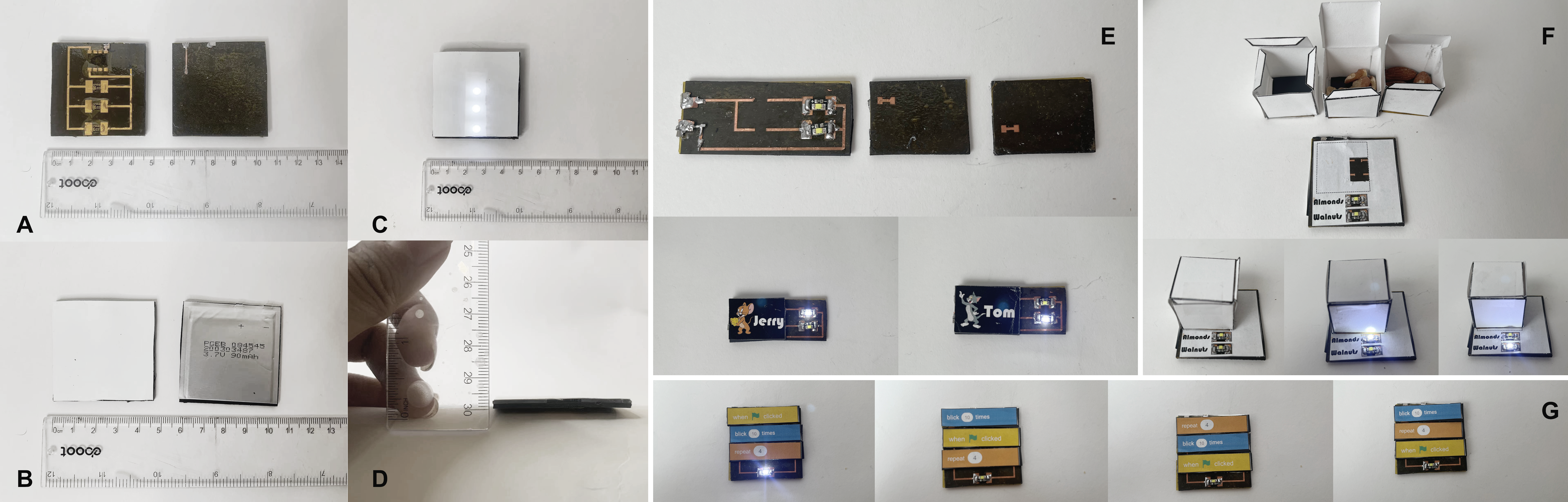}
  \caption{Electronic magnetic sheet: (A). Exposed PCBs of sheets A and B (B) Display and battery sides. (C) When the complementary sheets A and B combine, the LEDs light up and blink every second. (D) The thickness of the electronic magnetic sheet is 3mm. (E) Application: Authentication User Interface. Two PCBs, here labeled Tom and Jerry, will selectively pair in different configurations on the larger bottom PCB, lighting up different LEDs with Tom and Jerry IDs. (F) Application: Object classification. Three magnetically-folded boxes of equal size contain different contents. A classifier that couples the magnetic interactions to a force transducer communicates the contents of the boxes via LEDs without opening them. (G) Tangible Interface for learning the Scratch language. Scratch components will only bond in locations where they form correct semantic meanings, and light up an LED when placed there to communicate this to the student.}
  \label{fig: Applications}
\end{figure}

\textbf{Testing}: Once each individual laminate is assembled, we superimpose the two sheets in their correct mating configuration and confirm magnetic attraction and a strong electrical connection (Figure \ref{fig: Applications}A,B,C,D). The sheets will not bond readily in any other configuration, helping the user find the correct orientation and avoiding shorting the circuit.


\section{Applications}

We illustrate use cases of our composite material and fabrication process through applications in physical authentication, classification, tangible interfaces and foldable structures. 

\subsection{Double-authentication user interface}
We can leverage the selectivity of the magnetic codes together with the closing of the electrical circuit for aligned PCBs as a double-authentication method, as both magnetic bonding and electrical connection are required to verify correct bonding. Using PCBs populated with LEDs to communicate this authentication visually, we can enrich the feedback to the user in a way that is tailored to the application (Figure \ref{fig: Applications}E).

\subsection{Object detection, classification and tracking}
Object detection, classification, and tracking are critical features to logistics, transportation, and warehousing management, and ways to do so unobtrusively are particularly salient in HCI. With our technique, we leverage magnetic codes to build a sheet that bonds to itself when folded to form a cube. We fill the cubes with contents before folding the last side, it's lid. Outward-facing magnetic codes on each cube allow each cubic "package" to be identified and sorted without opening it (Figure \ref{fig: Applications}F).

\subsection{Tangible interface for learning Scratch for K12 student}
Scratch is a visual programming language aimed primarily for children as an educational tool for programming. Using our rapid fabrication technique, we can physically instantiate these visual programs as interactive physical coding blocks for students to learn and play with. These blocks can be magnetically coded to only bond to create a particular "program", and designed electrically to visually confirm correct "compilation" using LEDs (Figure \ref{fig: Applications}G).

\subsection{Interactive, foldable structures}
\par
\begin{figure}
  \includegraphics[width=\linewidth]{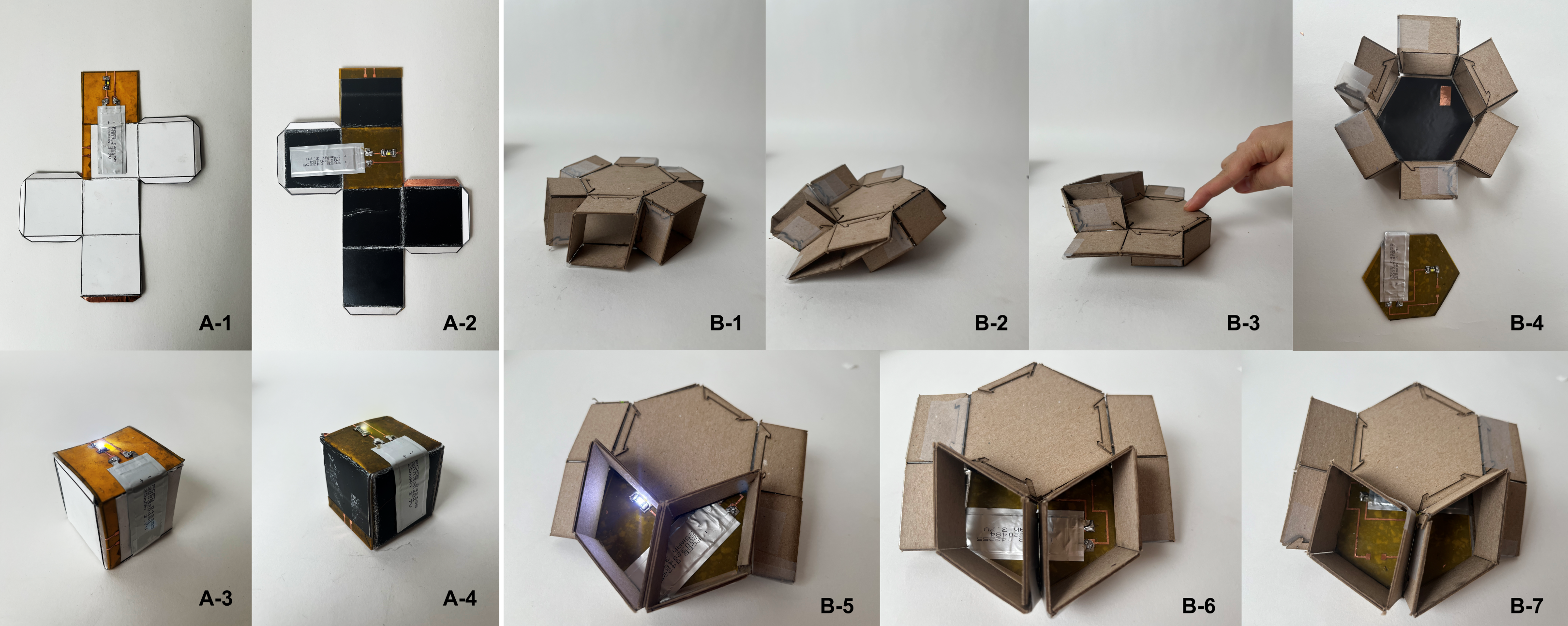}
  \caption{Conveying successful deployment of folded structures. (A) An unfolded cube can be folded in two directions to create either a black or white cube, with different LEDs signaling correct target acquisition. (B) This mechanism can similarly be used in multistable structures to signal that a particular geometry has been acquired by the object.}
  \label{fig: Apps-origami}
\end{figure}

 
Folding is an efficient method to develop bistable and dynamic surfaces~\cite{huang2012easigami}. Using our technique, we fabricate 2D sheets and fold them into 3D geometries, embedding circuits onto sides of what becomes different folded solids (Figure \ref{fig: Apps-origami}). In Figure \ref{fig: Apps-origami}A, An unfolded cube is colored with white on one side and black on the other. It can be folded in either direction to create either a white or black cube as seen by the user, and a different LED will light up to confirm correct folding on completion. In Figure \ref{fig: Apps-origami}B, a more exotic shape has multiple stable configurations, and the LED only lights up on folding into the particular desired shape.

\section{Limitations and future work}
\par
In this section, we discuss the limitations of our fabrication process and propose directions for future work.


\subsection{Fabrication streamlining}
Our existing fabrication procedure involves three separate user interfaces and three different machines, resulting in several disaparate steps in both the design and fabrication procedures. To unburden user attention and decrease the risk for human error, future work will streamline file conversion and material alignment between these procedures.

\subsection{Magnetic sheet thickness}
Increasing the magnetic sheet thickness increases the bonding force available between mating pairs. While our design trade-off weighed inexpensive, commercially available alternatives against each other to select our sheet, materials exhibiting higher flux densities, such as high-purity iron particulate cast in silicone matrices, could be explored to raise bond strength while retaining low material thicknesses.

\subsection{Integrating origami mechanisms}
Origami is a traditional Japanese art that has received widespread used in engineering~\cite{peraza2014origami}. It has been shown that single single sheets can be constructed to acquire multiple different target configurations once folded in order to adapt the sheet to diverse or changing needs~\cite{reis2015transforming}. Future work should explore leveraging these capabilities using our approach to develop more exotic applications in self-folding structures.


\section{Conclusion}
\par
In this paper, we introduced a computational composite sheet comprised of actuation, energy and computation layers and developed a fabrication process for its construction. We leveraged inexpensive materials and readily available fabrication machines to support the rapid exploration of applications from computational composites. We illustrated how this sheet can be used to build new applications for physical authentication, classification, tangible interfaces and foldable structures. In particular, at a sheet price of \$6 per 50mm square, it supports the inexpensive creation of novel interactive devices that can be assembled rapidly and are capable of communicating its state to the user via onboard LEDs. Finally, we surveyed the limitations to our approach and highlighted avenues for future work for continuing the improvement of portable, computational composite materials for interactive applications.

\bibliographystyle{ACM-Reference-Format}
\bibliography{sample-base}

\end{document}